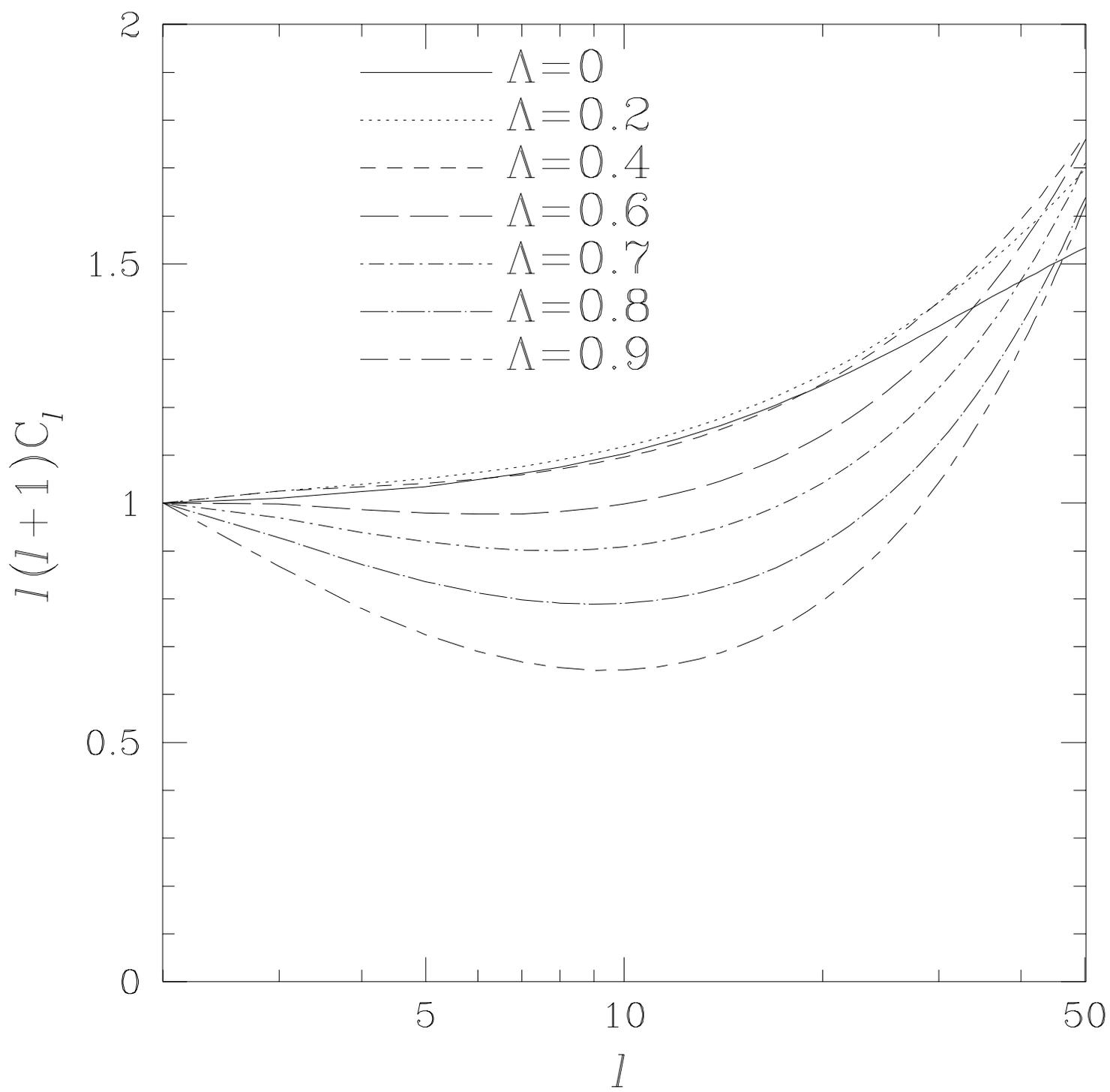

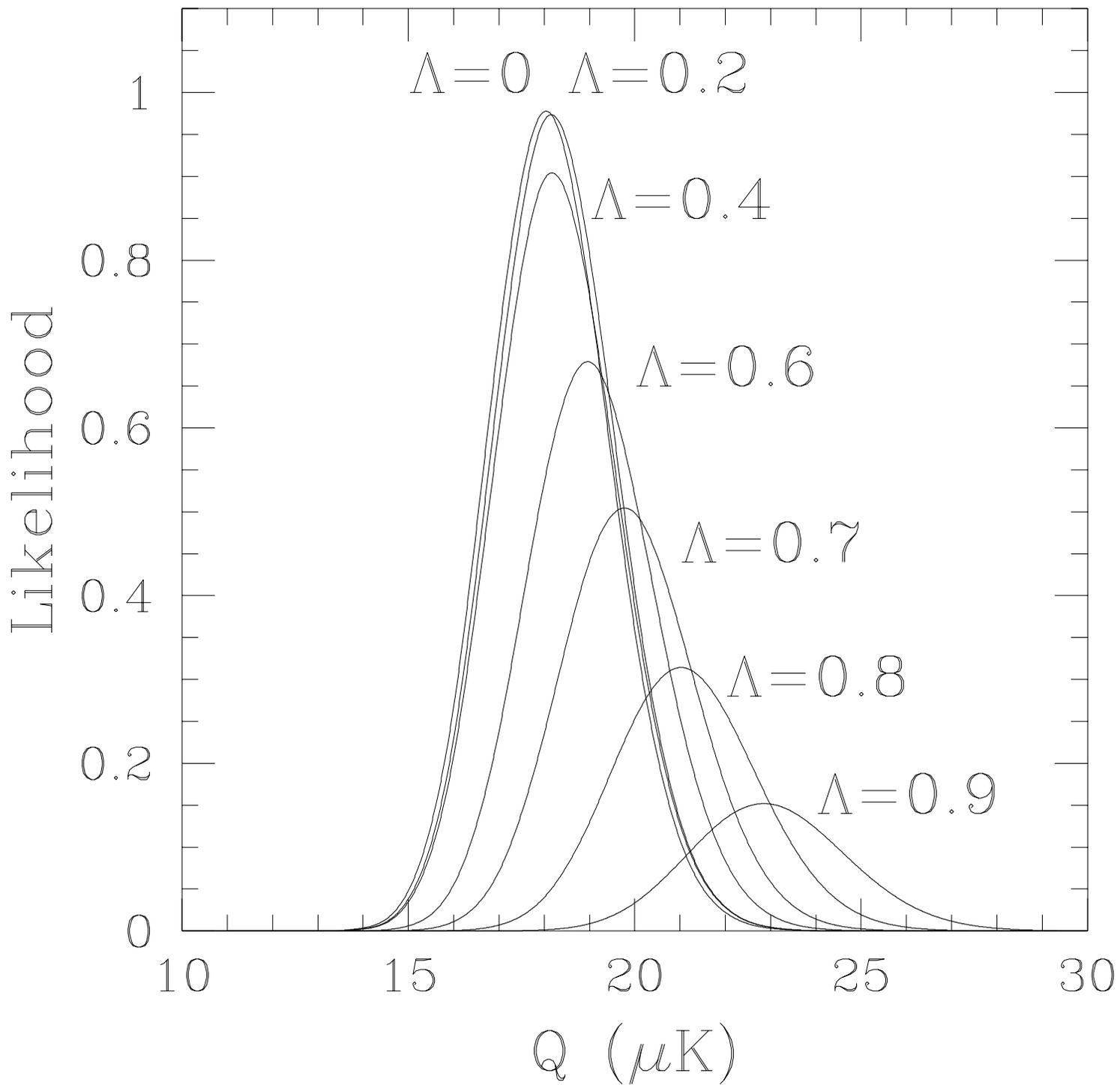

# Cosmological-Constant Cold Dark Matter Models and the *COBE* Two-Year Sky Maps[†]

Emory F. Bunn[1] and Naoshi Sugiyama[1,2]

[1]*Departments of Physics and Astronomy and Center for Particle Astrophysics*
*University of California, Berkeley, CA 94720*

[2]*Department of Physics, Faculty of Science*
*University of Tokyo, Tokyo 113, Japan*

*E-mail: bunn@pac2.berkeley.edu, sugiyama@pac3.berkeley.edu*

**Abstract.**

We compare the two-year *COBE* DMR sky maps with the predictions of cosmological-constant cold dark matter ($\Lambda$CDM) models. Using a Bayesian analysis, we find that the most likely value of the cosmological constant in a $\Lambda$CDM model is $\Lambda = 0$. The data set an upper limit on $\Lambda$ of 0.78 (0.85) at 90% confidence, and 0.86 (0.92) at 95% confidence with (without) the quadrupole anisotropy. The angular power spectrum $C_l$ depends only quadratically on $\Lambda$ when $\Lambda$ is small; the likelihood function $L(\Lambda)$ is therefore quite flat near $\Lambda = 0$.

Subject headings: cosmic microwave background — large-scale structure of universe

# Introduction.

Over the last few years, both the quantity and the quality of cosmological data have improved rapidly. Several new observations and data analyses of the large-scale structure of the Universe have been performed (see e.g., Peacock & Dodds 1994). Moreover, the recent detection of anisotropy in the cosmic microwave background (CMB) by the *COBE* satellite (Smoot *et al.* 1992) provides important information about the initial conditions of the fluctuations. It is now possible to test models for the growth of large-scale structure in the Universe more precisely than ever before. The general picture that appears most consistent with these data is that structure formed by gravitational instability from small perturbations in the early Universe. Within this paradigm, however, there are many unanswered questions. The manner in which structure evolves depends upon the amount and kind of matter in the Universe, as well as the statistical properties of the initial fluctuations. The most popular model so far is the cold dark matter (CDM) model, in which the Universe is assumed to be cosmologically flat, with density parameter $\Omega = 1$. A small fraction (around a few percent) of the matter in this model is baryonic, while the rest is nonrelativistic collisionless matter of some kind. The initial perturbations are assumed to be Gaussian and adiabatic, with a power-law power spectrum $P(k) = k^n$, with $n \approx 1$.

The CDM model's popularity arises from the fact that it is approximately consistent with observations of galaxy clustering, and with observations of the CMB anisotropy. However, as the quality of these data has improved, including the detection of the CMB anisotropy by the COBE DMR (Smoot *et al.* 1992), significant discrepancies have arisen between the observed power spectrum of galaxy clustering and the CDM prediction (Efstathiou, Bond, & White 1992, Peacock & Dodds 1994). In order to resolve this problem, there have been a variety of proposed modifications to the CDM model: among the proposed fixes are tilting the primordial power spectrum (see e.g., Davis *et al.* 1992), mixing some amount of hot dark matter with the cold (see e.g., Davis, Summers & Schlegel 1992) and assuming a small value for the Hubble constant (Bartlett *et al.* 1994). One very promising modification involves replacing some of the cold dark matter with a cosmological constant $\Lambda$ while keeping a flat geometry (see e.g., Efstathiou, Sutherland & Maddox 1990; Efstathiou, Bond & White 1992; Kofman, Gnedin & Bahcall 1993). This has the added advantage of solving the cosmic age problem.

It is known, however, that $\Lambda$ models produce quite different CMB anisotropies on large scales (Kofman & Starobinski 1985), because of the time evolution of the gravitational potential which produces the Sachs-Wolfe integral (Sachs & Wolfe 1967). Sugiyama & Silk (1994) numerically solved the perturbation equations and obtained CMB power spectra for different $\Lambda$'s. The power spectra for $\Lambda$-dominated models have shapes that are quite different from the simple power-law models that are usually used in analyzing the CMB anisotropy. Therefore, these power-law analyses (e.g., Smoot *et al.* 1992, Seljak & Bertschinger 1993, Górski *et al.* 1994) cannot be used to set limits on $\Lambda$. A direct comparison between the DMR data and the predicted power spectra from the Λ models is required.

In this paper we compare predicted CMB power spectrum of ΛCDM models with the *COBE* DMR data. In these models, the Universe is supposed to be cosmologically flat, so that the density parameter $\Omega$ and the cosmological constant $\Lambda$ add up to unity. The matter in the Universe is assumed to consist primarily of CDM, with a few percent of the matter being baryonic. It should be noticed that the CMB anisotropy on the angular scales probed by *COBE* is largely independent of the matter content of the Universe. Therefore our results obtained here are general for ΛCDM models. The initial fluctuations are supposed to be Gaussian and adiabatic, with a Harrison-Zel'dovich ($n = 1$) power spectrum, as in standard CDM. These models fit the galaxy clustering data quite well with $\Lambda \sim 0.8$ (Efstathiou, Sutherland, & Maddox 1990; Efstathiou, Bond, & White 1992; Kofman, Gnedin & Bahcall 1993).

## Method.

We compare the DMR data with the ΛCDM predictions using a likelihood technique similar to that described by Bond (1994a, 1994b). We will assume that the CMB anisotropy seen by the DMR has the properties of a Gaussian random field. It is useful to write the anisotropy as an expansion in spherical harmonics:

$$\frac{\Delta T}{T}(\hat{\mathbf{r}}) = \sum_{l,m} a_{lm} Y_{lm}(\hat{\mathbf{r}}). \tag{1}$$

(Here and hereafter, all sums over $l$ run from 2 to $\infty$, and all sums over $m$ run from $-l$ to $l$. Modes with $l = 0, 1$ are marginalized over during the likelihood analysis.) The coefficients $a_{lm}$ are independent Gaussian random variables with zero mean and variances

$$C_l \equiv \left\langle |a_{lm}|^2 \right\rangle, \tag{2}$$

where the angle brackets denote an ensemble average. The power spectrum $C_l$ is related to the rms temperature fluctuation in the following way:

$$\left\langle |\Delta T/T|^2 \right\rangle = \sum (2l+1) C_l / 4\pi. \tag{3}$$

The angular power spectrum $C_l$ is determined by the cosmological model under consideration. Our goal is to determine which models (*i.e.*, which angular power spectra) are consistent with the DMR data.

The DMR data consist of $N_p = 6144$ pixels. After excising all pixels within 20° of the Galactic plane, the number of pixels remaining is $N = 4038$ (for maps that are pixelized in ecliptic coordinates). Each pixel contains a measurement of $\Delta T/T$ at a point $\hat{\mathbf{r}}_i$, smoothed with the DMR beam pattern, and contaminated with noise: the $i$th data point can be represented as

$$d_i = \left(\frac{\Delta T}{T} \star B\right)(\hat{\mathbf{r}}_i) + n_i = \sum_{l,m} B_l a_{lm} Y_{lm}(\hat{\mathbf{r}}_i) + n_i, \qquad (4)$$

where $B$ represents the beam pattern and the star denotes a convolution. $B_l$ is the $l$th coefficient of the expansion of the beam pattern in Legendre polynomials. It can be approximately modeled as a Gaussian, $B_l \approx \exp(-0.5l(l+1)/19.3^2)$, but we use the more accurate values from Wright *et al.* (1994). $n_i$ is the noise in the $i$th pixel. To a good approximation, the noise is uncorrelated from pixel to pixel (Lineweaver *et al.* 1994), and the quantities $n_i$ can be accurately modeled by Gaussian random variables with known variances $\sigma_i^2$.

Our goal is to assess the consistency of various cosmological models with the data $d_i$. Ideally, we would like to compute the likelihood of each model,

$$-2\ln L = \vec{d}^T M^{-1} \vec{d} + \ln\det M, \qquad (5)$$

where the covariance matrix $M$ is determined by the angular power spectrum $C_l$ predicted by the model:

$$M_{ij} = \langle d_i d_j \rangle = \sum_{l,m} B_l^2 C_l Y_{lm}(\hat{\mathbf{r}}_i) Y_{lm}^*(\hat{\mathbf{r}}_j) + \sigma_i^2 \delta_{ij}. \qquad (6)$$

Unfortunately, it is impractical to compute $L$ for each model we wish to test: $4038 \times 4038$ matrices are not easy to invert. We are therefore led to consider ways of approximating the likelihood function.

Since the DMR beam pattern drops off rapidly with $l$, we expect that a relatively small number of modes, say those with $l < l_{max}$, contribute significantly to the spherical harmonic expansion of $\Delta T/T$. It therefore seems plausible that we could "compress" the data vector in some judicious way to a dimension $D \sim l_{max}^2$ that would be significantly smaller than $N$. Then the likelihoods would be much easier to compute. If we performed this compression in the right way, we could hope that the compressed data would contain most of the useful information in the raw data. If the data covered the sky completely, then the obvious way to effect this compression would be to use the data to estimate the individual spherical harmonic coefficients: $b_{lm} = (4\pi/N) \sum_{i=1}^{N} d_i Y_{lm}(\hat{\mathbf{r}}_i)$. In the case of incomplete sky coverage, the orthogonality of the spherical harmonics is destroyed, and this procedure no longer provides reliable estimates of the true coefficients (Peebles 1980; Bunn, Hoffman & Silk 1994). However it is still perfectly acceptable to use these quantities for computing likelihoods, as long as one takes proper account of their covariance matrix (Górski 1994).

We will follow a slightly different approach. In order to avoid repeated inversion of large matrices, we will compress the data in the following way before computing likelihoods. We will choose some $D \times N$ matrix $A$ with $D < N$, and compute $\vec{u} = A\vec{d}$. Then for any model we wish to test, we can compute the $D \times D$ covariance matrix of $\vec{u}$, $M_u = AMA^T$, and the likelihood,

$$-2 \ln L_u = \vec{u}^T M_u^{-1} \vec{u} + \ln \det M_u. \tag{7}$$

In order for these likelihoods to be useful in constraining models, we need to make a good choice for the matrix $A$. In particular, we would like to choose $A$ in such a way that the likelihood function $L_u$ has maximum rejection power. Suppose the correct power spectrum is $C_l$. Then if we make a small change to the power spectrum, say by changing the normalization by a factor $(1 + \epsilon)$, we would like the likelihood to decrease as much as possible on average. That is, we would like to minimize $\langle \delta L_u \rangle \equiv \langle L_u(C_l) - L_u((1+\epsilon)C_l) \rangle$. Since $\langle L_u \rangle$ has a maximum at the true model $\epsilon = 0$, $\langle \delta L_u \rangle \propto \langle L_u'' \rangle$, the second derivative of $\langle L_u \rangle$ with respect to $\epsilon$ evaluated at the true model $\epsilon = 0$.

It is straightforward to calculate $\langle L_u'' \rangle$, and to minimize this quantity with respect to the matrix $A$. The optimal matrix $A$ turns out to be a matrix whose rows are solutions $\vec{a}$ of the generalized eigenvalue problem

$$M_{sig}\vec{a} = \lambda M \vec{a}, \tag{8}$$

where the covariance matrix $M$ consists of a signal part and a noise part: $M_{sig,ij} = \sum_{l,m} C_l B_l^2 Y_{lm}(\hat{\mathbf{r}}_i) Y_{lm}^*(\hat{\mathbf{r}}_j)$ and $M_{noise,ij} = \sigma_i^2 \delta_{ij}$. Furthermore, the eigenvectors we want are those with the largest eigenvalues $\lambda$.

This result is not difficult to interpret. The rows of $A$ are, to use Bond's terminology (Bond 1994a, 1994b), "eigenmodes of the signal-to-noise ratio." The components of the reduced data vector $u$ are simply the projections of the raw data vector $d$ in the directions of these eigenmodes. The modes with large eigenvalues are the directions in the $N$-dimensional data space which have large sensitivity to the signal, as compared with the noise. In this sense, the signal-to-noise eigenmodes are the optimal directions to look at in data space.

Of course, in order to determine the eigenmodes, we need to make a guess about the actual power spectrum of the data. However, it turns out that there is little danger in guessing wrong: the likelihood function $L_u$ does not change significantly as the input power spectrum is varied. For definiteness, from now on we will take the rows of $A$ to be the solutions of (8) with a Harrison-Zel'dovich power spectrum.

We also need to decide on a value for $D$, the dimension of the reduced data vector $\vec{u}$. We want $D$ to be large enough to have good rejection power, but small enough to be computationally convenient. We have found by trial and error that the likelihood functions for $D = 400$ and $D = 500$ are almost identical, whereas the $D = 300$ likelihood function is noticeably broader, signifying less rejection power. We have therefore chosen to use $D = 400$ in testing our models. With $D = 400$, it is possible to compute all rows of $A$ by truncating all sums over $l$ at $l_{max} = 30$: increasing $l_{max}$ to 40 leaves these eigenmodes unchanged.

Although we remove a best-fit monopole and dipole from the data, in the absence of complete sky coverage we cannot be sure we have removed precisely the correct monopole and dipole. We therefore marginalize over these quantities by integrating over all possible values of the monopole and dipole. This is the correct procedure when performing a Bayesian analysis; an alternative procedure would be to take the peak likelihood as the monopole and dipole are allowed to range over all possible values; this procedure gives results which are negligibly different from marginalizing.

### Results.

We have applied the technique presented in the previous section to the *COBE* DMR two-year sky maps. The DMR data consist of six sky maps, two independent channels at each of three frequencies. We formed one sky map by taking a linear combination of all six maps, with weights proportional to the inverse square of the noise levels in the maps. We excised all pixels within 20° of the Galactic plane, and removed a best-fit monopole and dipole from the data. We then computed the signal-to-noise eigenmode matrix $A$ and the reduced data vector $\vec{u}$ for this data set. It is now straightforward to compute the likelihood $L_u$ for any model we wish to test.

Before examining $\Lambda$CDM models, we tested our method on standard power-law power spectra. In an $\Omega = 1$ cosmology, on angular scales large enough that the only significant source of anisotropy is the Sachs-Wolfe effect (Sachs & Wolfe 1967), the angular power spectrum is

$$C_l = \left(\frac{4\pi Q^2}{5}\right) \frac{\Gamma\left(\frac{2l+n-1}{2}\right)\Gamma\left(\frac{9-n}{2}\right)}{\Gamma\left(\frac{2l+5-n}{2}\right)\Gamma\left(\frac{3+n}{2}\right)}, \qquad (9)$$

where the power spectrum of density perturbations is $P(k) \propto k^n$ and the quadrupole $Q \equiv Q_{rms-PS} \equiv \sqrt{5C_2/4\pi}$ is used to normalize the power spectrum (Abbott & Schaefer 1986; Bond & Efstathiou 1987). We computed the likelihood function $L_u(n, Q)$ and used it to constrain the slope $n$ and normalization $Q$ of the power spectrum. We found that the maximum-likelihood solution was $(n, Q) = (1.3, 17.1\,\mu\text{K})$. Furthermore, using standard Bayesian techniques with a uniform prior, we found that $0.64 < n < 1.78$ at 95% confidence, and that $17.9 < Q < 24.3$ for $n = 1$ at 95% confidence. Both results are consistent with previous analyses (Górski *et al.* 1994).

Figure 1 shows the angular power spectra $C_l$ for seven $\Lambda$CDM models. These power spectra are obtained by direct numerical calculations up to the present epoch (see Sugiyama & Gouda 1992). All have Hubble constant $H_0 = 50\,\text{km}\,\text{s}^{-1}\,\text{Mpc}^{-1}$ and baryon density $\Omega_B = 0.03$. The cosmological constant $\Lambda$ takes the values $0, 0.2, 0.4, 0.6, 0.7, 0.8, 0.9$. (The model with $\Lambda = 0$ is, of course, standard CDM.) The dependence on $\Omega_B$ of $C_l$ is quite weak, i.e., less than 10% for $l \lesssim 30$. The dependence on the Hubble constant is also weak: changing $H_0$ from 50 to 80 leaves the $C_l$'s almost identical for $l \lesssim 10$, and causes less than a 20% difference for $l \lesssim 30$. It is clear from Figure 1 that the $C_l$'s for these models are quite different from flat or simple power-law models, particularly when $\Lambda$ is large. For each of these models, we computed $L_u$ as a function of the power spectrum normalization $Q$. These likelihoods are shown in Figure 2. It is clear that the data prefer low values of $\Lambda$.

We can place Bayesian confidence limits on $\Lambda$ in the following way. For each of the seven models, we compute the marginal likelihood by integrating over the normalization $Q$:

$$L_{marg}(\Lambda) \equiv \int L_u(\Lambda, Q) \, dQ. \tag{10}$$

These marginal likelihoods are plotted in Figure 3 as a function of $\Lambda$. We then smoothly interpolate between the data points, using a natural cubic spline. We can then say that $\Lambda$ has an upper bound $\Lambda_{max}$ at some confidence level $c$ if

$$\int_0^{\Lambda_{max}} L_{marg}(\Lambda) \, d\Lambda = c \int_0^1 L_{marg}(\Lambda) \, d\Lambda. \tag{11}$$

The upper limit on $\Lambda$ derived in this way is, to use the standard statistical nomenclature, the boundary of a Bayesian credible region in the $Q$-$\Lambda$ plane, adopting a prior distribution that is uniform in both $Q$ and $\Lambda$ (Berger 1985). A uniform distribution is the proper "noninformative prior" for $\Lambda$. Several different choices of prior distribution could be justified for $Q$; however, since the data contain a very strong detection, the results are insensitive to the prior in $Q$ (Bunn et al. 1994). Table 1 shows the upper limits for $\Lambda$ for several different confidence levels.

An alternative procedure would be to use the maximum value of the likelihood, $L_{max}(\Lambda) \equiv \max_Q L_u(\Lambda, Q)$, in place of the marginal likelihood in eq. (11). In practice, it makes very little difference which procedure is followed: replacing $L_{marg}$ by $L_{max}$ changes the 95% confidence level upper limit on $\Lambda$ from 0.86 to 0.87.

The quadrupole moment of the DMR sky maps is anomalously low, and it is possible that it may be contaminated in some way (see e.g., Górski et al. 1994). We explored the sensitivity of our results to this possibility by removing the quadrupole from the data and repeating the analysis including only modes with $l \geq 3$, marginalizing over the quadrupole as well as the monopole and dipole. Table 1 contains the upper limits on $\Lambda$ found in this way. Since the large quadrupole predicted by $\Lambda$CDM models is at odds with the low quadrupole in the data, it is not surprising that removing the quadrupole weakens the constraints.

The likelihood curve in Figure 3 is flat near $\Lambda = 0$. The reason for this is easily seen by looking at Figure 1: the angular power spectrum for $\Lambda = 0.2$ is almost identical to the $\Lambda = 0$ model, indicating that $C_l$ is a very weak function of $\Lambda$ for small $\Lambda$. Specifically, if we perform a Taylor expansion of $C_l$ about $\Lambda = 0$, the linear term is negligibly small.

## Conclusions.

We find that ΛCDM models are strongly disfavored by the DMR data. In particular, models with $\Lambda > 0.86$ are inconsistent with the data at a confidence level of 95%, while models with $\Lambda > 0.78$ are inconsistent at the 90% level. The large-scale structure data seem to prefer ΛCDM models with $\Lambda \gtrsim 0.7$. It is clear from these results that the DMR data are at best marginally consistent with such models.

Removing the quadrupole from the data weakens these constraints somewhat. However, it is important to note that the primary reason for removing the quadrupole is simply that it is anomalously low. Removing data points simply because they do not fit our theoretical expectations is a dangerous statistical practice. In the absence of strong independent evidence that the quadrupole is contaminated, one should be wary of throwing it away.

These results we obtained here are about the same as the best current limits from gravitational lensing (Kochanek 1993, 1994). It should be noticed that our method is totally independent and free from any ambiguities, such as the evolution and mass distribution of galaxies, involved in the gravitational lensing method. There is every reason to expect that, once the full four-year DMR data have been analyzed, it will be possible to discriminate between models with $\Lambda \approx 0.6$ and those with $\Lambda = 0$ with very high confidence, since the power spectra in these models have very different shapes (see Figure 1).

## Acknowledgements.

We would like to thank D. Scott, J. Silk, M. Tegmark and W. Hu for useful discussions. N.S. acknowledges financial support from a JSPS postdoctoral fellowship for research abroad.

**Table 1.** Upper limits on $\Lambda$

| Confidence level | $\Lambda_{max}$ including $Q$ | $\Lambda_{max}$ excluding $Q$ |
|---|---|---|
| 68% | 0.55 | 0.62 |
| 90% | 0.78 | 0.85 |
| 95% | 0.86 | 0.92 |
| 99% | 0.96 | 0.98 |

**Figure Captions.**

1. The angular power spectrum $l(l+1)C_l$ is shown for each of the seven $\Lambda$CDM models described in the text. The power spectra have been normalized so that $6C_2 = 1$.

2. The likelihood $L_u$ is plotted as a function of $Q$ for each of the seven $\Lambda$CDM models described in the text. The overall normalization is arbitrary.

3. The marginal likelihood $L_{marg}$ is plotted as a function of $\Lambda$. The solid curve is the likelihood including the quadrupole, and the dashed curve is the result of excluding the quadrupole. The crosses show the marginal likelihood for each of the seven models described in the text. The smooth curves are cubic spline interpolations between them. The overall normalization is arbitrary.

# References.

Abbott, L., & Schaefer, R.K., Ap. J., 308, 546 (1986).

Bartlett, J.G., Blanchard, A., Silk, J., & Turner, M.S., 1994, submitted to Nature.

Berger, J.O. 1985, *Statistical Decision Theory and Bayesian Analysis*, Springer-Verlag, New York.

Bond, J.R. 1994a, Astrophys. Lett. and Comm., in press.

Bond, J.R. 1994b, preprint (CITA–94-27, astro-ph–9407044).

Bond, J.R. & Efstathiou, G. 1987, M.N.R.A.S., 226, 655.

Bunn, E.F., Hoffman, Y., & Silk, J. 1994, Ap. J., 425, 359.

Bunn, E.F., White, M., Srednick, M., & Scott, D. 1994, Ap. J., 429,1.

Davis, M., Summers, F.J., & Schlegel, D., 1992, Nature, 359,393.

Davis, R.L., Hodges, H.M., Smoot, G.F., Steinhardt, P.J., & Turner, M.S., 1992, Phys. Rev. Lett., 69, 1856.

Efstathiou, G., Sutherland, W.J., & Maddox, S.J., 1990, Nature, 348, 705.

Efstathiou, G., Bond, J.R., & White, S.D.M., 1992, M.N.R.A.S., 258, 1p.

Górski, K. 1994, preprint (*COBE* preprint 94–07, astro-ph–9403066).

Górski, K.,*et al.* 1994, preprint (*COBE* preprint 94–08, astro-ph–9403067).

Kochanek, C.S. 1993, Ap. J., 419, 12.

Kochanek, C.S. 1994, in *Critique of the Sources of Dark Matter in the Universe*, in press.

Kofman, L.A. & Starobinski, A.A. 1985, Sov.Astron.Lett., 11, 271.

Kofman, L., Gnedin, N. & Bahcall, N. 1993, Ap. J., 413, 1.

Lineweaver, C. *et al.* 1994, preprint (astro-ph–9403021).

Peacock & Dodds 1994, M.N.R.A.S., 267, 1020.

Peebles, P.J.E. 1980, *The Large-Scale Structure of the Universe*, Princeton University Press.

Sachs, R.K. & Wolfe, A.M. 1967, Ap. J., 147, 73.

Seljak, U. & Bertschinger, E. 1993, Ap. J., 417, L9.

Sugiyama, N. & Gouda, N. 1992, Prog. Theor. Phys., 88, 803.

Sugiyama, N. & Silk, J. 1994, Phys. Rev. Lett., 73, 509.

Smoot *et al.* 1992, Ap. J., 396, L1.

Wright, *et al.* 1994, Ap. J., 420,1.

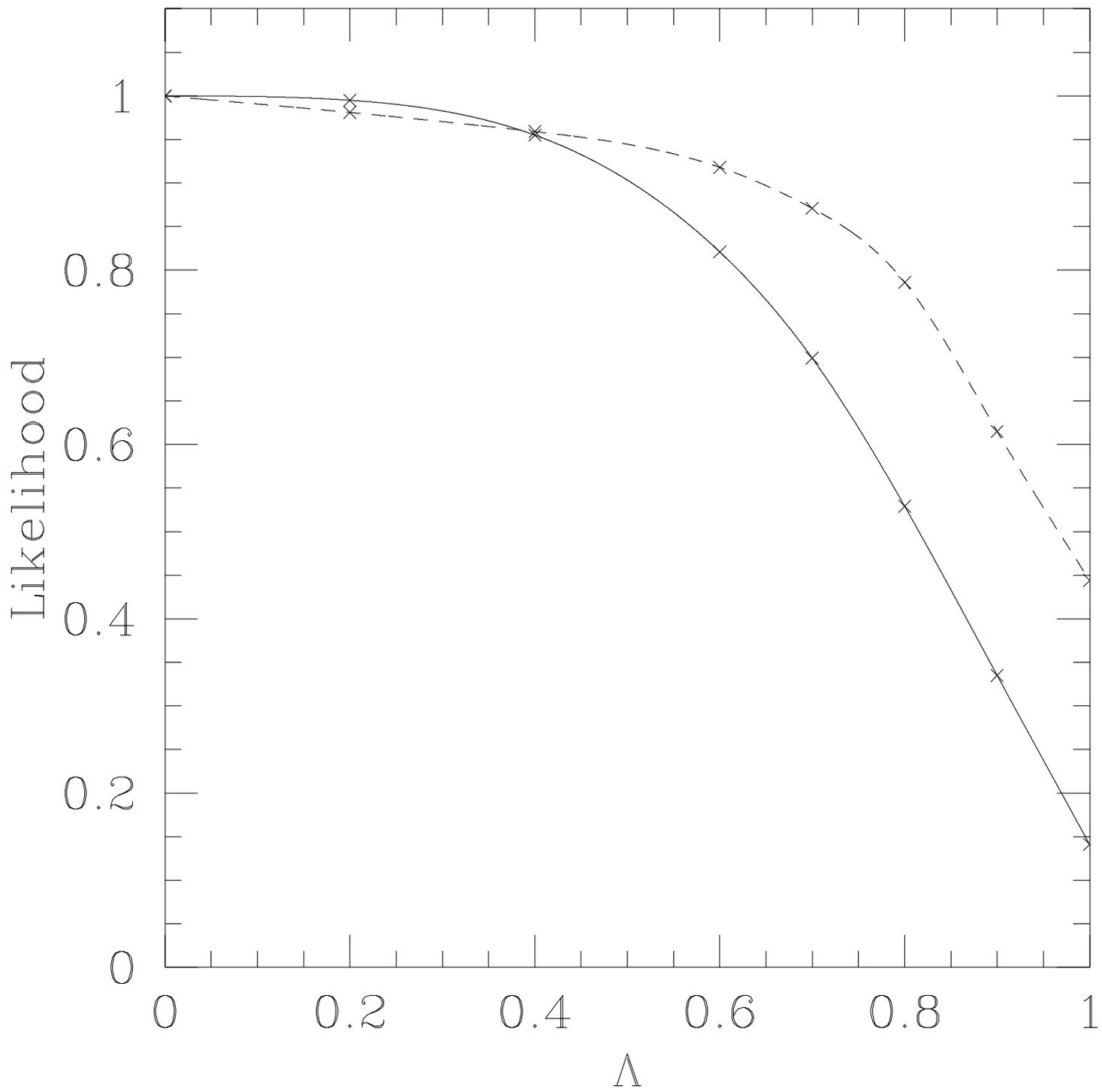